\documentclass{article}

\usepackage{arxiv}

\usepackage[utf8]{inputenc} 
\usepackage[T1]{fontenc}    
\usepackage{hyperref}       
\usepackage{url}            
\usepackage{booktabs}       
\usepackage{amsfonts}       
\usepackage{nicefrac}       
\usepackage{microtype}      
\usepackage{graphicx}
\usepackage{natbib}
\usepackage{doi}

\title{Implementing engrams from a machine learning perspective: matching for prediction.}

\author{ \href{https://orcid.org/0000-0001-7914-8494}{\includegraphics[scale=0.06]{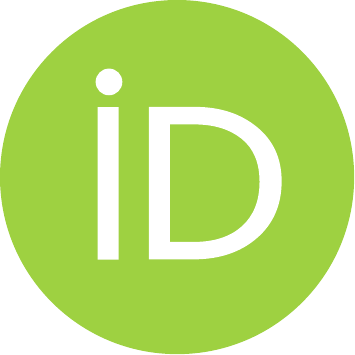}\hspace{1mm}Jesus Marco de Lucas} \\
	Advanced Computing and e-Science Group\\
	Instituto de Fisica de Cantabria (IFCA) CSIC-Universidad de Cantabria\\
	Santander, ES 39005 SPAIN\\
	\texttt{jesus.marco@csic.es} \\
}

\renewcommand{\headeright}{Short Comment}
\renewcommand{\undertitle}{Short Comment}
\renewcommand{\shorttitle}{\textit{arXiv} Implementing Engrams}

\hypersetup{
pdftitle={Implementing Engrams},
pdfsubject={q-bio.NC, q-bio.QM},
pdfauthor={Jesus.~Marco},
pdfkeywords={Engrams, Autoencoders, Prediction by Matching},
}

\begin{document}
\maketitle

\begin{abstract}
Despite evidence for the existence of engrams as memory support structures in our brains, there is no consensus framework in neuroscience as to what their physical implementation might be.

Here we propose how we might design a computer system to implement engrams using neural networks, with the main aim of exploring new ideas using machine learning techniques, guided by challenges in neuroscience.

Building on autoencoders, we propose latent neural spaces as indexes for storing and retrieving information in a compressed format. We consider this technique as a first step towards predictive learning: autoencoders are designed to compare reconstructed information with the original information received, providing a kind of predictive ability, which is an attractive evolutionary argument.

We then consider how different states in latent neural spaces corresponding to different types of sensory input could be linked by synchronous activation, providing the basis for a sparse implementation of memory using concept neurons.

Finally, we list some of the challenges and questions that link neuroscience and data science and that could have implications for both fields, and conclude that a more interdisciplinary approach is needed, as many scientists have already suggested.

\end{abstract}


\section{Introduction}
Neuroscience is probably the most challenging field of research today, given the intrinsic complexity and diversity of biological structures in the brain, the impact of advances in this field on our current and future lives, and the many challenges that are being addressed (Herrera et al., 2020). 

Our brain achieves incredible levels of performance compared to our current human-designed, electronic-based computers, both in terms of capabilities and energy consumption, so it seems natural to try to find inspiring challenges from the study of our brain to develop new ideas in computer science, even if in many cases these challenges in neuroscience will not be solved without new experimental breakthroughs.

In fact the study of our brain has been a source of inspiration for artificial intelligence (AI) since its inception (Turing, 1950). Many advances in machine learning, from perceptrons (McCulloch and Pitts, 1943) to deep learning techniques (LeCun et al., 2015), have been inspired by the analogy with the structure of neural networks in our brains, even if we don't really know how they might handle learning tasks.

It must also be said that the exploration of some of the most successful ideas in machine learning to establish an analogy with neural processes has been pursued extensively, but technical problems, such as the difficulty of establishing backpropagation in neural circuits, have not allowed these solutions to be applied by direct analogy to understanding our brain (Lillicrap et al., 2020).

At present, both the neuroscientific and the machine learning scientific community promote the interest of this interplay at a global level (Richards et al., 2019; Zador et al., 2022).

One such inspiring global challenge is to understand how our brains store and retrieve information, i.e. how our memory processes work. The "engram", a term proposed by Richard Semon (Semon, 1921) to refer to the physical substrate of our memory, is still a very active topic of research. After long studies trying to find the localisation of "engrams" in the brain (Eichenbaum, 2016), and despite significant advances in the knowledge of neural mechanisms in recent years, the reality is that we do not know the details of how our brain stores the memories it perceives (Josselyn and Tonegawa, 2020; Berlot, Popp and Diedrichsen, 2018; Gebicke-Haerter, 2014; Han et al., 2022; Fuentes-Ramos, Alaiz-Noya and Barco, 2021).

In the field of machine learning, the issue of memory is considered somewhat secondary, as information is naturally stored digitally. However, the common interest in episodic memory, which is key for predictive tasks, and its relationship to attentional mechanisms (Vaswani et al., 2017) has significantly increased the potential convergence of the two fields in recent years.

As a starting point, we have chosen a fascinating question related to this challenge: the possible existence of "concept cells" (Quiroga, 2012). Concept cells are individual neurons that selectively fire at an image or text that corresponds to a given identity, as measured in the brains of different people in many different examples. The opinion paper "No Pattern Separation in the Human Hippocampus" (Quian Quiroga, 2020) summarises the very interesting chain of developments in this field, triggered by his team's discovery in 2005 of the initially referred as the "Jennifer Aniston" neurons, the first example of an idea that was previously proposed within the community without a clear scientific basis, the "grandmother cells" (Gross, 2002). The results were popularised by the media, as the images presented included celebrities such as Jennifer Aniston (JA in what follows), Halle Berry or Jackie Chan, to name but a few. While the arguments presented in the cited paper and the long list of accompanying references provide clear support for the basic hypothesis, i.e. that concept cells underlie the engrams that support conceptual memory, there is no explicit "mechanistic" model of how these engrams could be encoded in a "sparse" mode while at the same time leading to the activation of a single neuron.

\section{An analogy for structures supporting engrams: autoencoders}

Encoders, and in particular autoencoders ('Autoencoder', 2022), are a very interesting method that has benefited from the application of deep learning and convolutional neural network techniques. The basic idea is simple: to achieve a large reduction in dimensionality by applying a coding filter to large samples of data, such as images, and projecting them onto a multi-dimensional vector, the latent space.

Our approach is also simple, we start with the scheme of an engram (memory) system for storing images, which includes four different interrelated parts:

-an encoding system, which receives as input images from a vision system

-A latent space, a set of nodes that store the vector values that are the output of the encoding system.

-A decoding system capable of "recovering" the input image from a vector value in the latent space.

-Critically, a layer of "concept nodes" that connect all value points in the latent space that are related to the same concept.

\begin{figure}
	\centering
 \includegraphics[width=8cm]{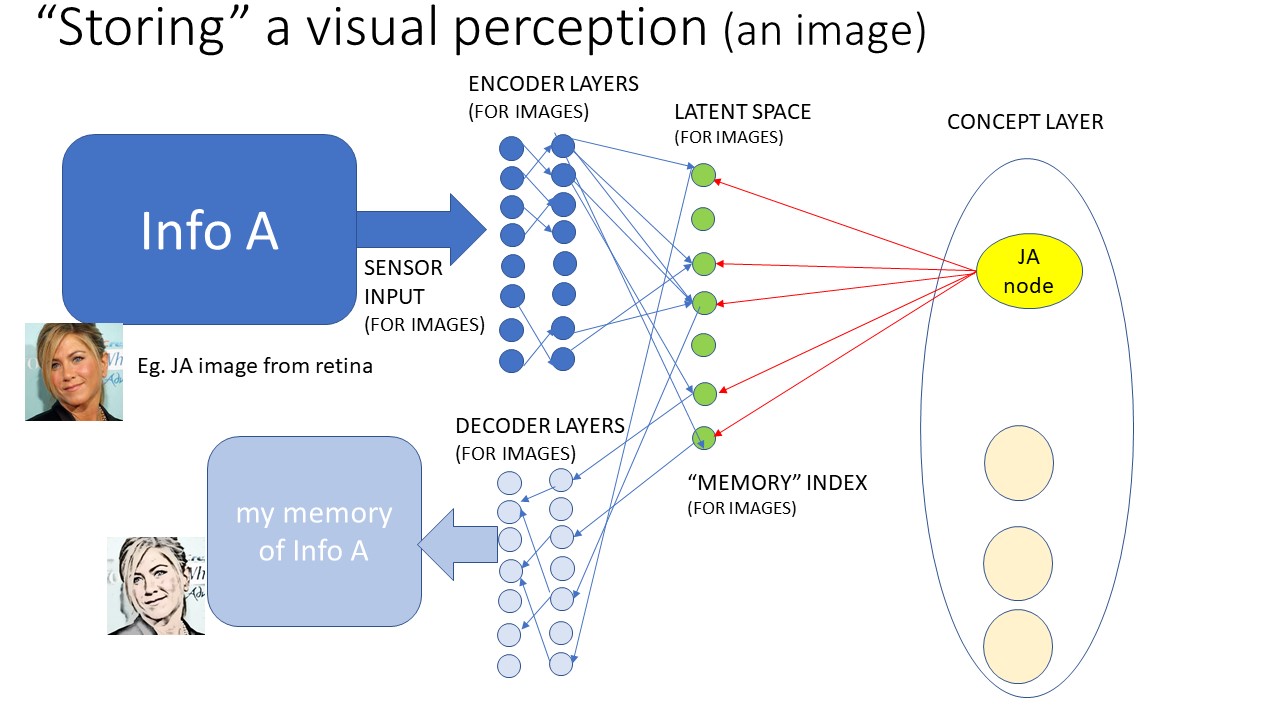}
 \includegraphics[width=8cm]{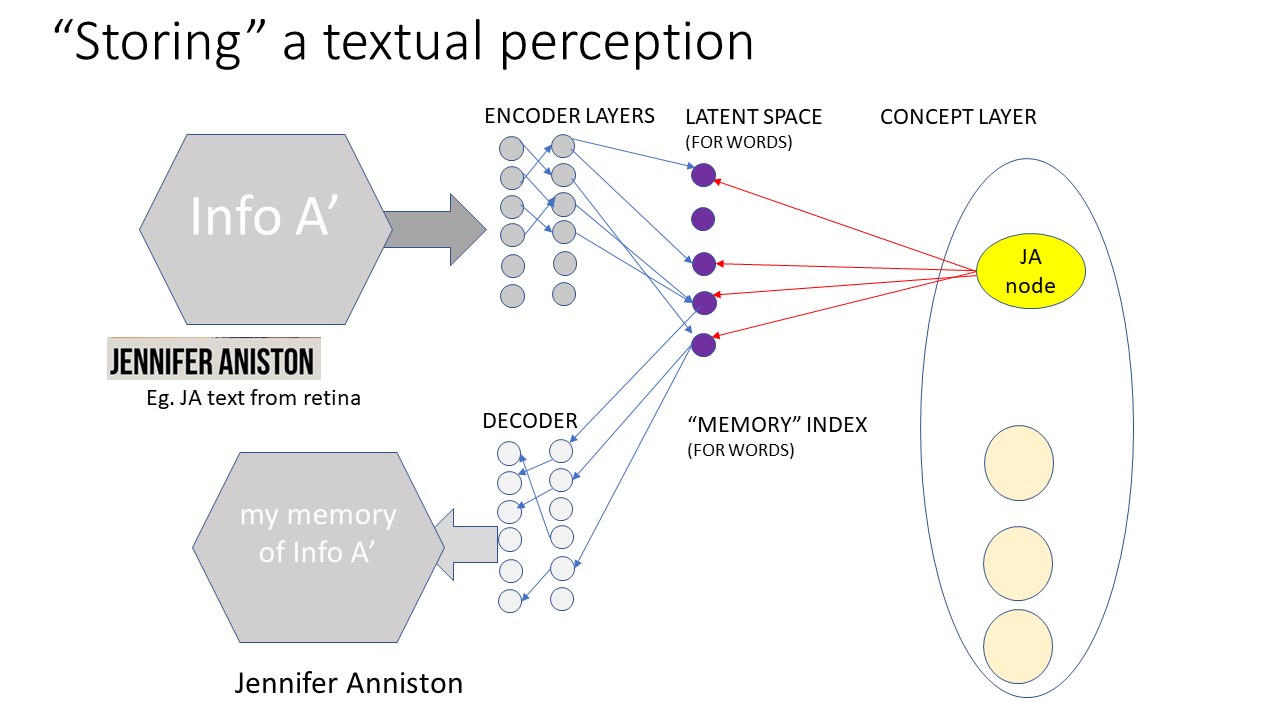}
 \includegraphics[width=16cm]{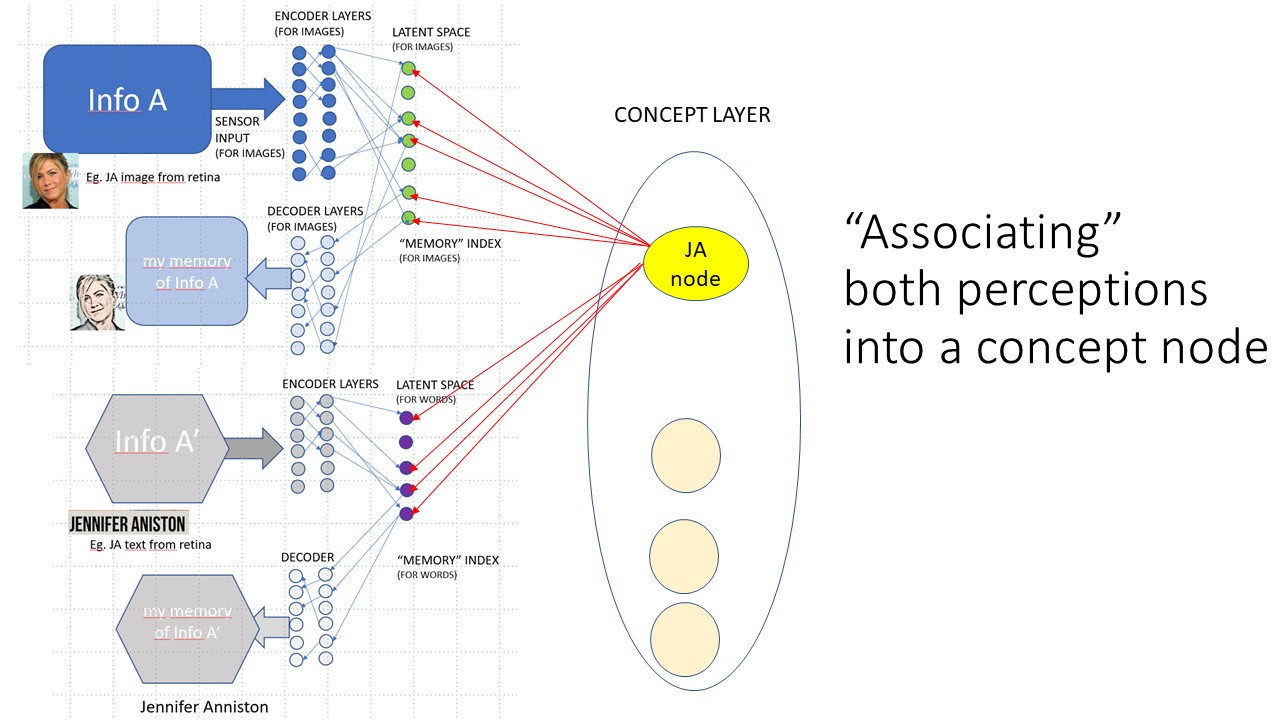}
	\caption{Basic scheme for storing information using auto-encoders in concept nodes. Note that the structure of the encoder/decoder system could be different for each type of information, defining different latent spaces. The identity of a concept, that is the concept node, is defined by the associations (values) of this concept node with the various vector values that the information corresponding to this concept receives in these latent spaces (pictures, texts, sounds, internal signals, etc.).}
	\label{fig:fig1}
\end{figure}

A basic didactic scheme of these ideas is shown in figure\ref{fig:fig1}. 

The previous scheme, which can be easily implemented in a computer, already raises interesting questions about a possible analogous implementation in the neuronal circuits of the brain.

This memory system could be built with neurons, dendrites or even sets of dendritic spines as nodes, while connections would be established as synapses, which could be either binary (0/1, active/not active) or "analogical" valued connections.

The first idea to be explored in this analogy is how the information might be encoded, and in particular what kind of loss function might make sense, and what level of compression might be appropriate.

As a first example, consider that our eyes as sensors have a resolution of about 500 Mpixels, while our visual memory capacity is probably much more limited, both as a short-term memory and, even more so, as a long-term memory. 

Thus, a first question to be raised following this analogy is what could be considered a realistic compression factor for images corresponding to a given person, and what kind of loss function makes sense. This question implies an analysis of the possible structure of the coding layer, the latent space, and also of the decoding layer. This decoding structure may or may not be symmetrical to the encoding structure, in which case the loss function may have a different resolution.

This question raises the very interesting point of what external information is perceived by our senses but lost when processed in our brain. And on the other hand, what artefacts might be created by our brain when it tries to recover the original information from the compressed information encoded in the latent spaces.

The second, much trickier point, which is also related to the first question, is how encoding might occur physically in our brains, and how we might improve our computational encoders by analogy. In our computers, we use deep learning techniques, including backpropagation and convolutional filters, to train these autoencoders, and we "force" the learning process to apply the desired loss function, for example by using a cross-entropy comparison. 

Here we have no answer, but some more didactic considerations. After all, the process of matching the original image with the reconstructed one can be seen as an example of "prediction", since recognition is the basis for further action. This is a very interesting evolutionary argument to support the idea that autoencoders could be implemented in our brains and, more generally, that engrams supporting memory have appeared as structures in living organisms. The idea is to look for examples where the basic mechanism, i.e. matching a sensory input with a neural feedback, is implemented and this configuration is selected by a biological system.

There is another related argument to consider regarding autoencoders and their possible implementation as neural networks in animals: given their "simple" structure and scalability, it is possible that encoders could be considered as a basic structure that could be genetically defined, providing newborns with memory capacity and basic learning functions without the need for initial training on a large data set. There is a clear analogy with transfer learning methods in machine learning, where the initial architecture and weights of a new neural network for a specific task are provided by a previous neural network trained on a very large database of images.

\section{Building a concept node: associating sparse memory over encoders}

In all nervous systems, including our brain, the external data is provided by perception, either images, including symbols, from vision, or sound, taste, smell, etc. There are also many internal data channels, including chemical and electrical signals.

To define a complete JA neuron, all the related information must be linked. What is more, this information usually needs to be linked to a wider context.

Returning to our computational scheme, we have a vector value in the latent space corresponding to a visual perception, a picture of JA, another vector value in the latent space corresponding to the text "Jennifer Aniston", and so on. The JA node simply links all these values together so that the complete concept JA can be recovered from these connections.

This structure can be implemented as a noSQL database, where each node corresponds to a concept.  For computers, this is an associative task, associating a key (label JA) with different values. 

For our neurons in the brain, we could speculate that the initial association of an image and a text, which will give rise to a label or key, could be triggered by synchronicity, following the well-known formula "neurons that fire together, wire together" (Hebb, 1949). The JA neuron would link the representations in the two different latent spaces of the image and the text, as they would fire at the same time. 

We can also imagine that when we see a new text with the words 'Jennifer Aniston', the vector value in the latent space for words will be the same, and in this way we can establish a link with the JA node.

However, it is not obvious how we can use the existing autoencoder to identify a new image of JA and link it to the existing key, since the new vector value in the latent space of images will not in principle be close to the previous vector value for the first image. We need a computational solution that applies a supervised classification on top of an unsupervised solution, which is what the autoencoder is.

There are several possible computational approaches to this problem. The first could be quite direct: we apply transfer learning starting from a pre-existing neural network that has been trained in supervised mode on a very large database of images that have been classified, and use the new images to refine these pre-existing categories. For example, ResNet (He et al., 2015) is a neural network that, once trained on the ImageNet database (Deng et al., 2009), we could use as a basis for transfer learning. Following this approach, we could modify the previous architecture and consider the latent space as a pre-classifier layer. A new image of JA would initially be given the same values to classify in the scheme (or thesaurus) used by the training database (in the previous case it would be WordNet (Miller, 1995) and the image would be classified as corresponding to a woman), and in this way it could be linked to the previous image, which already exists as a concept node. As new images are initially labelled, a supervised classification scheme begins to be defined.

A second approach, which is probably not too different in practice, could be to build the autoencoder in such a way that different images of the same concept correspond to very close points in the latent space, so that a concept is defined by locality in such a latent space, and the connection of one image to a previous one is defined by a minimum distance in this latent space. The structure of the latent space could be extended in this way to support hierarchical conceptualisation, providing a kind of generalisation. 
Several proposals, such as concept splatters (Grossmann, Gröller and Waldner, 2022), have already been made to structure latent spaces in this direction.

We can now begin to have an initial scheme for organising the different information associated with a given 'identity' such as JA. All visual information is "encoded" in a neural network oriented towards encoding images, and a corresponding latent space; all textual information is "encoded" in another neural network oriented towards encoding words, and a corresponding latent space; all information perceived as sound is "encoded" in another neural network, and so on. The "concept" neurons are implemented as indexes that connect the latent spaces and allow joint retrieval of the information associated with the "concept". The complete machine learning system to support engrams would be a combination of autoencoders embedded in deep convolutional neural network classifiers developed using transfer learning. The concept nodes would be indexes stored in a hierarchical noSQL database linking the points in the corresponding latent spaces.

Translating this scheme to our brain, we can consider the ideas described long ago (Teyler and DiScenna, 1986; Treves and Rolls, 1994) and further developed in several papers proposing computational models involving the hippocampus (Kesner and Rolls, 2015), which is usually considered a central hub for many cognitive activities, including memory (Lisman et al., 2017). 

In our inspiring model, concept neurons, and presumably latent space encoding, could be located in hippocampal areas, while the corresponding encoding-decoding neural networks that support most of the information in the engram could be developed in the different cortical areas already identified by their different functional properties: visual cortex, language area, etc. 

This proposal overcomes the apparent conflict of a limited storage capacity due to the limited number of neurons in the hippocampus, whose main function would be to support indexing activity and associative connections to configure a conceptual space or cognitive map, as also discussed in recent work (Whittington et al., 2022). These associative connections would be established by synchrony, between points in different latent spaces, and by locality, between different points in the same latent space.

\section{Global reflection and next questions}

Our ultimate goal is to learn from the knowledge and ideas of neuroscience to advance machine learning: we know that our brain is a much more energy-efficient machine than our computers, and that it is better at complex and abstract tasks.

The complexity of the brain is reflected in the diversity and increasing number of publications on different topics in neuroscience. Similarly, publications on data science methods and applications have increased exponentially in recent years, following their success in solving problems that can be considered as artificial intelligence questions.

In this sense, we have shown how a challenging question in neuroscience, the implementation of engrams in our brain, can trigger an interesting analysis from a computational point of view. Moreover, once a certain hypothesis in neuroscience is considered, whether correct or not, in this case the existence of concept neurons, different questions and possible solutions can be proposed from a data science point of view, stimulating the search for new ideas.

The technical and scientific complexity of both fields, neuroscience and data science, makes interdisciplinarity a must in order to advance both fields together, as demanded by both communities.

Developing such an exploration initially from only one perspective may generate many hypotheses, most of them wrong, but some of them may stimulate reflection from the other perspective. What we can be sure of is that the search for answers would benefit from more intense collaboration.

Note that the proposal we have presented is not very original from the point of view of machine learning, since both autoencoders and NoSQL databases are well-known solutions for storing information. The key question is whether we could improve the design or combination of these computational tools by knowing how our brain works.  

Following a bottom-up analysis in neuroscience, the first question is to better understand neurons as cells, and also many other cells in the brain, to be able to explore their individual and collective properties either in simulations or in nanoprototypes, to understand their functionality, and to try to integrate these features into computational neural networks. In this respect, there are new possibilities to be explored with respect to the architecture of autoencoders, following the recent knowledge of the brain connectome and the relevant role of inhibitory neurons (Shapson-Coe et al., 2021), astrocytes (Labate and Kayasandik, 2023) or the consideration of computation at the dendritic level (Acharya et al., 2022).

It's worth remembering, however, that we don't yet have a realistic simulation of a cell, and that neurons are a very complex and diverse type of cell. There are many topics being studied in the neuron as a cell, from metabolism to the origin of electrical potentials and excitability, even more at the dendritic level, and most of them could be crucial for understanding how neuronal circuits also work as complex systems.

In any case, our main interest would be to find an idea of how the brain could assemble these cells and process the internal signals to be able to learn and memorise in such an efficient and powerful mode compared to our current techniques in machine learning. 

From our point of view, the main conclusion of this short note is that it would be interesting to explore, from a neuroscientific point of view and probably from an evolutionary perspective, an energy-efficient biological mechanism providing almost instantaneous basic pattern-matching capabilities, similarly to what autoencoders do using time-consuming and energy-intensive machine learning methods.

\bibliographystyle{unsrtnat}
\bibliography{references}  

Acharya, J. et al. (2022) ‘Dendritic Computing: Branching Deeper into Machine Learning’, Neuroscience, 489, pp. 275–289.  https://doi.org/10.1016/j.neuroscience.2021.10.001.

‘Autoencoder’ (2022) Wikipedia. https://en.wikipedia.org/w/index.php?title=Autoencoder (Accessed: 15 January 2023).

Berlot, E., Popp, N.J. and Diedrichsen, J. (2018) ‘In search of the engram, 2017’, Current Opinion in Behavioral Sciences, 20, pp. 56–60. https://doi.org/10.1016/j.cobeha.2017.11.003.

Deng J., Dong W., Socher R., L. -J. Li, Kai Li and Li Fei-Fei, "ImageNet: A large-scale hierarchical image database," 2009 IEEE Conference on Computer Vision and Pattern Recognition, Miami, FL, USA, 2009, pp. 248-255, https://doi.org/10.1109/CVPR.2009.5206848. 

Eichenbaum, H. (2016) ‘Still searching for the engram’, Learning and Behavior, 44(3), pp. 209–222. https://doi.org/10.3758/s13420-016-0218-1.

Fuentes-Ramos, M., Alaiz-Noya, M. and Barco, A. (2021) ‘Transcriptome and epigenome analysis of engram cells: Next-generation sequencing technologies in memory research’, Neuroscience \& Biobehavioral Reviews, 127, pp. 865–875. https://doi.org/10.1016/j.neubiorev.2021.06.010.

Gebicke-Haerter, P. (2014) ‘Engram Formation in Psychiatric Disorders’, Frontiers in neuroscience, 8, p. 118. https://doi.org/10.3389/fnins.2014.00118.

Gross, C.G. (2002) ‘Genealogy of the “Grandmother Cell”’, The Neuroscientist, 8(5), pp. 512–518. https://doi.org/10.1177/107385802237175.

Grossmann, N., Gröller, E. and Waldner, M. (2022) ‘Concept splatters: Exploration of latent spaces based on human interpretable concepts’, Computers \& Graphics, 105, pp. 73–84.  https://doi.org/10.1016/j.cag.2022.04.013.

Han, D.H. et al. (2022) ‘The essence of the engram: Cellular or synaptic?’, Seminars in Cell \& Developmental Biology, 125, pp. 122–135. https://doi.org/10.1016/j.semcdb.2021.05.033.

He, K. et al. (2015) ‘Deep Residual Learning for Image Recognition’. arXiv. https://doi.org/10.48550/arXiv.1512.03385.
Hebb, D.O. (1949) : A Neuropsychological Theory. New York: Psychology Press (2002).  https://doi.org/10.4324/9781410612403.

Herrera, E. et al. (2020) ‘Brain, Mind and Behaviour’. Libros Blancos. Desafíos Científicos 2030 del CSIC, vol. 5 https://doi.org/10.20350/digitalCSIC/12652.

Josselyn, S.A. and Tonegawa, S. (2020) ‘Memory engrams: Recalling the past and imagining the future’, Science, 367(6473), p. eaaw4325. https://doi.org/10.1126/science.aaw4325.

Kesner, R.P. and Rolls, E.T. (2015) ‘A computational theory of hippocampal function, and tests of the theory: New developments’, Neuroscience \& Biobehavioral Reviews, 48, pp. 92–147. https://doi.org/10.1016/j.neubiorev.2014.11.009.

Kim, D. et al. (2016) ‘Synaptic competition in the lateral amygdala and the stimulus specificity of conditioned fear: a biophysical modeling study’, Brain Structure and Function, 221(4), pp. 2163–2182. https://doi.org/10.1007/s00429-015-1037-4.

Labate, D. and Kayasandik, C. (2023) ‘Advances in quantitative analysis of astrocytes using machine learning’, Neural Regeneration Research, 18(2), p. 313. https://doi.org/10.4103/1673-5374.346474.

LeCun, Y., Bengio, Y. and Hinton, G. (2015) ‘Deep learning’, Nature, 521(7553), pp. 436–444. Available at: https://doi.org/10.1038/nature14539.

Lillicrap, T.P. et al. (2020) ‘Backpropagation and the brain’, Nature Reviews Neuroscience, 21(6), pp. 335–346. https://doi.org/10.1038/s41583-020-0277-3.

Lisman, J. et al. (2017) ‘Viewpoints: how the hippocampus contributes to memory, navigation and cognition’, Nature Neuroscience, 20(11), pp. 1434–1447. https://doi.org/10.1038/nn.4661.

McCulloch, W.S. and Pitts, W. (1943) ‘A logical calculus of the ideas immanent in nervous activity’, The bulletin of mathematical biophysics, 5(4), pp. 115–133. https://doi.org/10.1007/BF02478259.

Miller, George A. (1995). ‘WordNet: A Lexical Database for English’. Communications of the ACM Vol. 38, No. 11: 39-41. https://dl.acm.org/doi/10.1145/219717.219748 

Quian Quiroga, R. (2020) ‘No Pattern Separation in the Human Hippocampus’, Trends in Cognitive Sciences, 24(12), pp. 994–1007. https://doi.org/10.1016/j.tics.2020.09.012.

Quiroga, R.Q. (2012) ‘Concept cells: the building blocks of declarative memory functions’, Nature Reviews Neuroscience, 13(8), pp. 587–597. https://doi.org/10.1038/nrn3251.

Richards, B.A. et al. (2019) ‘A deep learning framework for neuroscience’, Nature Neuroscience, 22(11), pp. 1761–1770. https://doi.org/10.1038/s41593-019-0520-2.

Semon, R.W. (1921) The mneme. London: Allen\$ Unwin. http://archive.org/details/cu31924100387210

Shapson-Coe, A. et al. (2021) ‘A connectomic study of a petascale fragment of human cerebral cortex’. bioRxiv, p. 2021.05.29.446289. https://doi.org/10.1101/2021.05.29.446289.

Teyler, T.J. and DiScenna, P. (1986) ‘The hippocampal memory indexing theory’, Behavioral Neuroscience, 100, pp. 147–154.  https://doi.org/10.1037/0735-7044.100.2.147.

Treves, A. and Rolls, E.T. (1994) ‘Computational analysis of the role of the hippocampus in memory’, Hippocampus, 4(3), pp. 374–391. https://doi.org/10.1002/hipo.450040319.

Turing, A.M. (1950) ‘I.—COMPUTING MACHINERY AND INTELLIGENCE’, Mind, LIX(236), pp. 433–460. https://doi.org/10.1093/mind/LIX.236.433. 

Vaswani, A. et al. (2017) ‘Attention Is All You Need’, CoRR, abs/1706.03762. http://arxiv.org/abs/1706.03762

Whittington, J.C.R. et al. (2022) ‘How to build a cognitive map’, Nature Neuroscience, 25(10), pp. 1257–1272.  https://doi.org/10.1038/s41593-022-01153-y.

Whittington, J.C.R., Warren, J. and Behrens, T.E.J. (2022) ‘Relating transformers to models and neural representations of the hippocampal formation’. arXiv. https://doi.org/10.48550/arXiv.2112.04035.

Zador, A. et al. (2022) ‘Toward Next-Generation Artificial Intelligence: Catalyzing the NeuroAI Revolution’. arXiv. https://doi.org/10.48550/arXiv.2210.08340.







\end{document}